# 35.4 T field generated using a layer-wound superconducting coil made of (RE)Ba$_2$Cu$_3$O$_{7-x}$ (RE = Rare Earth) coated conductor

Ulf P. Trociewitz, Matthieu Dalban-Canassy, Muriel Hannion, David K. Hilton, Patrick Noyes, Youri Viouchkov, Jan Jaroszynski, Hubertus W. Weijers, and David C. Larbalestier

*National High Magnetic Field Laboratory, Tallahassee, FL, USA*

To explore the limits of layer wound (RE)Ba$_2$Cu$_3$O$_{7-x}$ (REBCO, RE = Rare Earth) coils in a high magnetic field environment > 30 T, a series of small insert coils have been built and characterized in background fields. One of the coils repeatedly reached 35.4 T using a single ~100 m length of REBCO tape wet wound with epoxy and nested in a 31 T background magnet. The coil was quenched safely several times without degradation. Contributing to the success of this coil was the introduction of a thin polyester film that surrounded the conductor. This approach introduces a weak circumferential plane in the coil pack that prevents conductor delamination that has caused degradation of several epoxy impregnated coils previously made by this and other groups.

The cuprate based high temperature superconductor (RE)Ba$_2$Cu$_3$O$_{7-x}$ (REBCO, RE = Rare Earth), has the capability to substantially transform the technology of high field magnet systems. So far, the low temperature superconductors Nb-Ti and Nb$_3$Sn have been used for virtually all superconducting high field magnets. Their maximum field, however, is limited by their upper critical fields ($H_{c2}$) of about 15 T for Nb-Ti and 30 T for Nb$_3$Sn, which limits their highest practical field to about 23.5 T[1]. This limit is imposed by the rapid decrease in critical current density $J_c$ as $H_{c2}$ is approached. By contrast, REBCO has an $H_{c2}$ that exceeds 100 T at 4.2 K, removing the $H_{c2}$ and $J_c$ limit that restricts usage of Nb$_3$Sn in high-field magnet systems. One of the goals at the NHMFL is to develop the necessary technology for the next generation of high-field magnets including Nuclear Magnetic Resonance (NMR) quality magnets. To reduce the number of resistive joints and achieve the required field homogeneity for NMR, layer-winding is highly desirable. Here we describe a series of four small test solenoids manufactured using REBCO coated conductor tape made by SuperPower Inc. All coils had an outer diameter of 38 mm and a winding inner diameter of 14 mm to fit into the bore of the 31 T resistive magnet at the National High Magnetic Field Laboratory (NHMFL) that provided the background field. The coils were instrumented with an array of voltage contacts to monitor consecutive sets of 5 - 10 layers throughout each winding pack. The specifications of these coils are shown in Table I. All coils were epoxy wet-wound to allow for homogeneous spatial load distribution exerted by the Lorentz forces during the high field runs. For two of the coils (A and B) 35 µm thick silk paper and a 0.4 mm diameter glass thread were used as interlayer and turn-to-turn insulation, respectively. This insulation, permeable to the low viscosity epoxy used for wet-winding, provided full bonding between the layers. Self-field critical current ($I_c$) measurements at 77 K, however, revealed severe damage of these coils that occurred in multiple layers without any apparent radial dependence. This and the moderate winding strain led us to look outside the winding process for the problem. Also, previous transport measurements on double-layer spirals with 14 mm inner diameter did not reveal any degradation on any of the monitored turns, including the conductor in-plane bend on the transition to the return layer where the conductor pitch angle changes. Moreover, other groups observed similar failure in fully epoxy-impregnated REBCO coils[2,3] and attributed the issue to transverse and peel stress concentration that caused internal conductor delamination within the buffer and the superconductor stack of layers at stresses as low as 0.5 MPa[4,5]. Peel stress in particular appears to be an issue in REBCO coated conductor. Polyanskii and Abraimov[6] observed incipient delamination along the edges of various as manufactured lengths of conductors. Evidence of delamination was also found in cross sections prepared from these coils, as shown in Figure 1. Our approach to alleviate this issue was to introduce an interface providing weak mechanical coupling between the conductor and epoxy by inserting the conductor into a thin wall, cryogenically compatible polyester shrink tube with a 12.7 µm initial wall thickness and close to 20 µm after shrinkage. The shrink tube was applied in sections of about 1.22 m with a 25 mm overlap between every section, which seals the conductor against seepage of the epoxy resin. The sections were then shrunk around the conductor using a temperature controlled heat source set at 150°C, which is about 50°C below the temperature at which we have observed conductor degradation to start. A short test coil (coil C) with 4 turns/layer and 62 layers was built with 25 m of REBCO conductor insulated this way and characterized at 4.2 K in a background field of up to 25 T. This coil generated an additional 2.7 T. The magnetic field dependence $I_c(H)$ of this coil is presented in Figure 1. At self-field, the $I_c$ of coil C is limited by the radial field of about 4 T that it generates at its ends. Comparing these data with short sample transport $I_c(H)$ and field orientation properties of conductor from the same batch[7,8] clearly showed that the coil fully reproduced short sample transport values and thus did not degrade. After the 4.2 K run, a thermal shock experiment was carried out by rapid cooling from RT to 77 K within seconds without causing any conductor degradation. We believe that the avoidance of degradation during initial cool-down, 4 K testing and severe thermal shock validates our hypothesis that the introduction of a weaker interface than that found within the conductor will protect epoxy-impregnated coils. Based on these results, a larger coil, labeled D shown in Figure 2, was built using the same approach. This coil was recently tested in a 31.2 T resistive magnet system at the NHMFL, generating an additional and steady magnetic field of 4.2 T for a total of 35.4 T at a bath temperature of 1.8 K. The specific field generation of the coil was 21.1 mT/A. Cooling the He bath to the superfluid region was useful due to the fact that He gas was trapped in the high $H \cdot \nabla H$ zone near the top end and terminal section of the coil, which slightly limited its performance and caused the measured $I_c$'s at highest fields to vary slightly[9]. Trapping of He gas was exacerbated by the fact that the coil almost entirely filled the 39 mm inner diameter of the cryostat leaving little space for cryogen. At full field, the coil operated at 196 A, which created significant maximum hoop stress levels of about 340 MPa, which however, were well within the maximum of about 760 MPa that we previously measured on REBCO layer-wound test coils[10]. At conductor dimensions of 4.02 mm x 0.096 mm, the engineering critical current density $J_e$ was > 500 A/mm². Quench protection of the coil was achieved by closely monitoring layer voltages, used to trigger a trip of the power supply and transfer the stored energy to a dump resistor circuit as soon as a threshold value was reached. With this safety circuitry, the coil was quenched safely multiple times without any measurable degradation. Analysis of the quench data revealed that the layer set 5–11 quenched first followed by layers 1–5 and 11–21 indicating that the quench originating in layers 5–11 propagated into the neighboring layer before driving the whole coil normal. The voltage evolution during the quench is shown in Figure 3. Layers 21–31, like the other remaining outer layers, showed a strong inductive response (rapid voltage drop on layers 21-31) but did not transition before the quench protection kicked in. Magnetometry data provided by the manufacturer give an indication as to why the coil quench was initiated at this location and not at the location of highest stress or most restrictive $I_c(H)$-field orientation. Taken at consecutive spots along the whole length of the conductor using a Hall sensor array moving along the conductor, those measurements showed one markedly lower magnetic moment within the conductor section wound in layers 5–11, where the quench initiated. Short length defects along REBCO coated conductors have been seen previously in microstructural and in magneto optical investigations on several samples and seem to be endemic at the present time[6]. This layer-wound coil and earlier prototype pancake-wound coils[10,11] demonstrate that REBCO coated conductors have developed into a suitable basis for a superconducting magnet technology that allows major advances in magnetic field generation at 4.2 K. Because of its low anisotropy and high irreversibility field at temperatures well above 50 K, REBCO conductor may also allow a cryo-cooled magnet technology for generating fields of 5 - 15 T in the 30 - 50 K range. A pancake-wound variant of this technology is currently being used for an all-superconducting 32 T user research magnet, which is now under construction at the NHMFL and is anticipated to be in service in 2013.

The authors thank Dima Abraimov, Valeria Braccini, Hom Kandel, Jun Lu, Anatolii Polyanskii, Michael Santos, and Andrew Whittington especially for their efforts in the characterization of the conductor used for the coils.

This work was supported by the NSF under DMR award 0654118.

Table I: Coil specifications.

| Coil | Inner Dia. (mm) | Outer Dia. (mm) | No. Turns (-) | No. Layers (-) | Conductor Length (m) |
|------|-----------------|-----------------|---------------|----------------|----------------------|
| A    | 14              | 38              | 22            | 60             | 110                  |
| B    | 14              | 38              | 22            | 70             | 112                  |
| C    | 14              | 38              | 4             | 62             | 22                   |
| D    | 14              | 38              | 15            | 80             | 96                   |

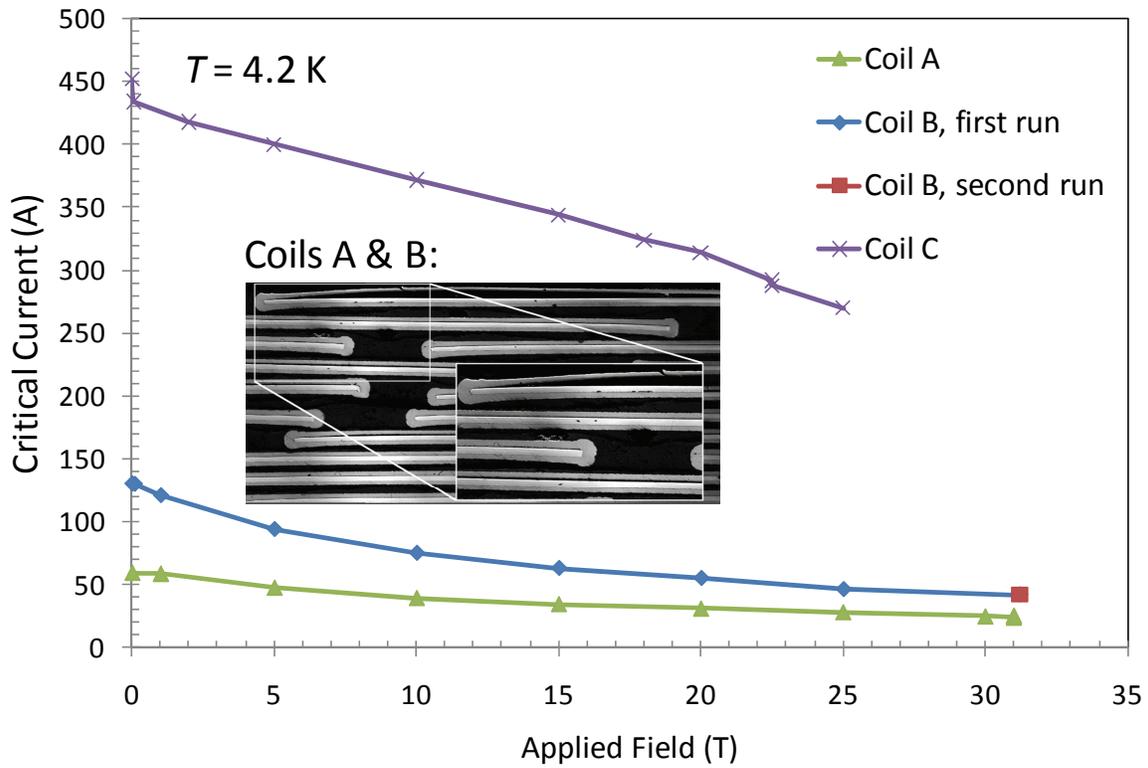

Figure 1: Critical current *vs.* applied field of coils A, B, and C showing the substantial difference in performance between fully epoxy impregnated coils and one that has been insulated with shrink tube. The inset is a cross sectional view of one of the earlier coils showing signs of conductor delamination at various locations throughout the winding pack.

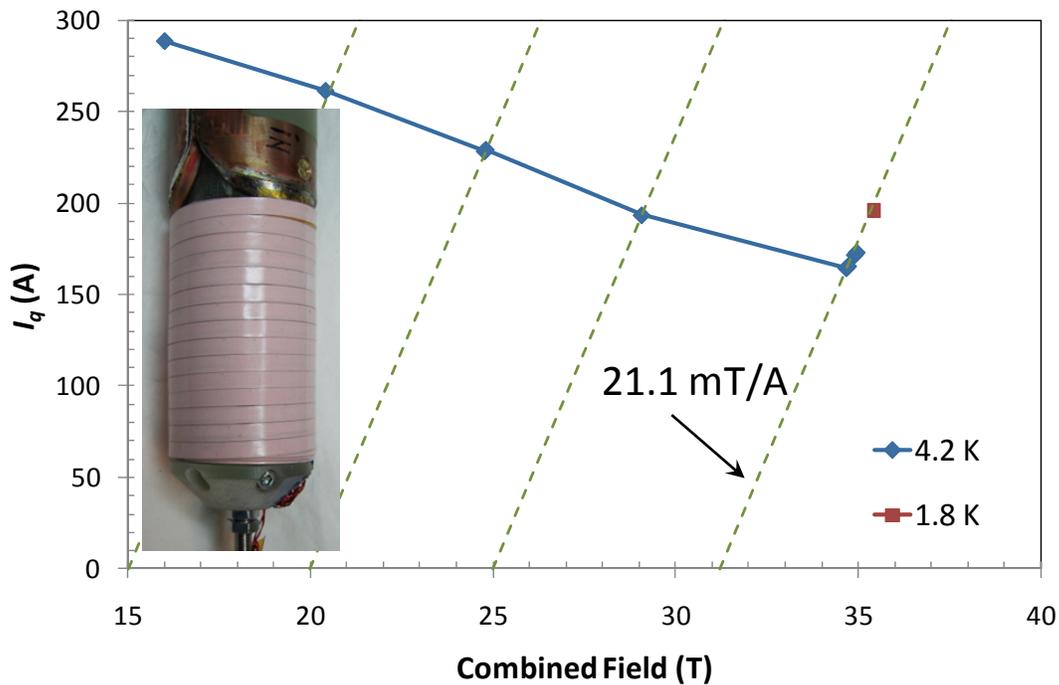

Figure 2: Quench currents *vs.* magnetic field for two bath temperatures: 4.2 K and 1.8 K. Heating developed close to the terminals due to helium gas trapping in that region, which reduced the quench currents in the field range above 20 T.

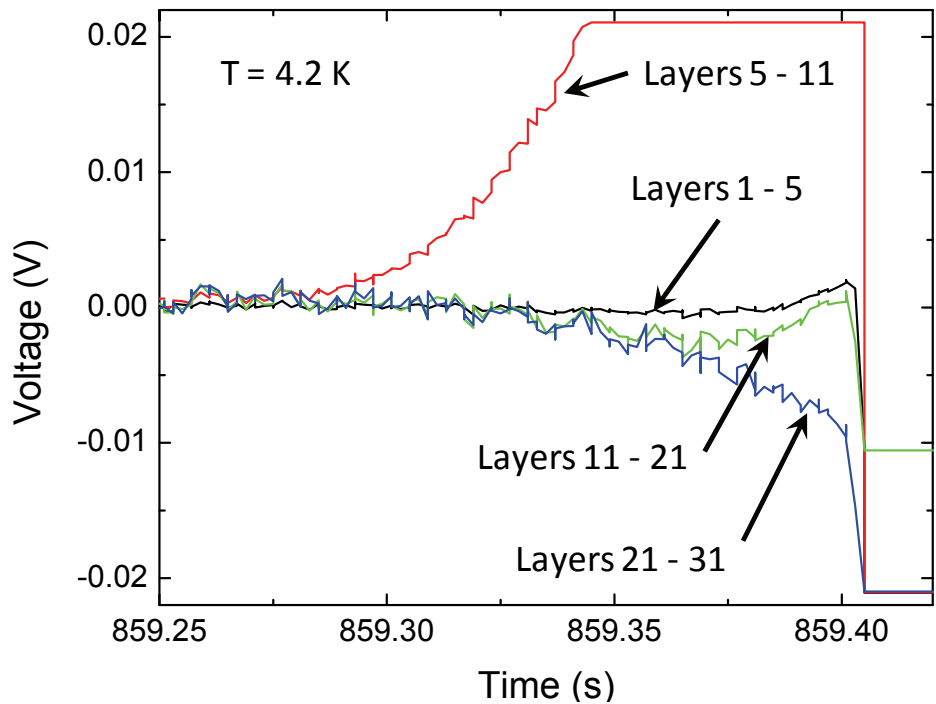

Figure 3: Voltage-time evolution during a quench of coil D. The layer set 5- 11 quenches first, driving the neighboring layers 1 – 5 and 11 – 21 normal soon after. The inset shows the actual coil. The leveling of the voltage at 859.35 ms is due to reaching the saturation of the signal amplifier used for the data acquisition. At 859.40 ms the protection system is triggered causing inductive voltage spikes that again are cut off by the saturation of the signal amplifier.